# Quantum critical behaviour of the plateau-insulator transition in the quantum Hall regime

A de Visser[1], L A Ponomarenko[1], G Galistu[1], D T N de Lang[1], A M M Pruisken[2], U Zeitler[3], D Maude[4]

[1]Van der Waals-Zeeman Institute, University of Amsterdam, Valckenierstraat 65, 1018 XE Amsterdam, The Netherlands

[2]Institute for Theoretical Physics, University of Amsterdam, Valckenierstraat 65, 1018 XE Amsterdam, The Netherlands

[3]High Field Magnet Laboratory, Radboud University, Toernooiveld 7, 6525 ED Nijmegen, The Netherlands

[4]Grenoble High Magnetic Field Laboratory, CNRS, 25 Avenue des Martyrs, Grenoble F 38042, France

E-mail: devisser@science.uva.nl

**Abstract**. High-field magnetotransport experiments provide an excellent tool to investigate the plateau-insulator phase transition in the integral quantum Hall effect. Here we review recent low-temperature high-field magnetotransport studies carried out on several InGaAs/InP heterostructures and an InGaAs/GaAs quantum well. We find that the longitudinal resistivity $\rho_{xx}$ near the critical filling factor $\nu_c \approx 0.5$ follows the universal scaling law $\rho_{xx}(\nu,T) \propto \exp(-\Delta\nu/(T/T_0)^\kappa)$, where $\Delta\nu = \nu-\nu_c$. The critical exponent $\kappa$ equals 0.56±0.02, which indicates that the plateau-insulator transition falls in a non-Fermi liquid universality class.

## 1. Introduction

The integral quantum Hall effect observed in semiconductor heterostructures is an excellent laboratory system to study quantum phase transitions (see Ref.1 and references therein). The plateau-plateau (PP) transitions present a sequence of quantum phase transitions, with the magnetic field, or filling factor, as control parameter. Upon further increasing the magnetic field the series of PP transitions is terminated by the plateau-insulator (PI) transition. Here the Fermi energy sweeps through the lowest Landau level and the transition is from the plateau $i = 1$ to the insulating phase. In recent years the PI transition has attracted much attention, because of its peculiar transport behaviour [2]. Low-temperature magnetotransport experiments in high-magnetic fields demonstrate that the longitudinal resistivity $\rho_{xx}$ diverges exponentially [3,4], whereas the Hall resistivity $\rho_{xy}$ remains quantized through the transition at the $i = 1$ plateau value, $\rho_{xy} = h/e^2$ [5,6]. At the critical filling factor $\nu_c \approx 0.5$, the $\rho_{xx}$ isotherms display a fixed point, with a critical resistivity $\rho_{xx,c} = h/e^2$. Consequently, the components of the conductivity tensor $\sigma_{xx}$ and $\sigma_{xy}$ follow the universal semi-circle relation $\sigma_{xx}^2+(\sigma_{xy}-e^2/2h)^2 = (e^2/2h)^2$, when the filling factor $\nu$ varies from 0 to 1 [7].

Recently, we have developed a methodology [1,8,9] to extract the critical behaviour (critical indices, scaling functions, flow diagrams) in the integral quantum Hall regime from the magnetotransport data. Especially, we have discovered that the PI transition is much more robust than the PP transitions against macroscopic sample inhomogeneities, such as contact misalignment and carrier density gradients. Therefore, we focus in our research on the plateau-insulator transition. In this paper, we briefly address the principles of scaling and delineate our methodology to disentangle the quantum critical aspects of the two-dimensional electron gas from the sample dependent effects. Next we review the results of recent high-field magnetotransport studies carried out on different InGaAs/InP heterostructures (HS) [1,6,8,10] and an InGaAs/GaAs quantum well (QW) [10,11].

## 2. Scaling functions and methodology

Let us write for the longitudinal and Hall resistivities of an ideal homogeneous sample $\rho_0$ and $\rho_H$, respectively. Within the scaling theory of the quantum Hall effect, at sufficiently low temperatures, these quantities with varying $B$ and $T$ become functions of a single scaling variable $X$ only (see e.g. Ref.1)

$$\rho_0(B,T) = \rho_0(X) \quad ; \quad \rho_H(B,T) = \rho_H(X) \qquad (1),$$

where
$$X = \frac{\nu - \nu_c}{\nu_0(T)} \quad ; \quad \nu_0(T) = (T/T_0)^\kappa. \qquad (2).$$

For the PI transition $\nu_c \approx 0.5$. The function $\nu_0(T)$ yields the universal critical exponent $\kappa$ and the sample dependent temperature scale $T_0$, which indicates the full width of the Landau level.

In general, the $\rho_{xx}$ and $\rho_{xy}$ data are affected by macroscopic sample inhomogeneities, which results in non-unique values when measured at different contact pairs of the Hall bar (see e.g. Ref.12). However, under simple circumstances (e.g. contact misalignment and carrier density gradients), the experimental resistivities are related to the ideal resistivities $\rho_0$ and $\rho_H$ by the relation [1]

$$\rho_{ij} = S_{ij}\rho_0(X) + \varepsilon_{ij}\rho_H(X) \qquad (3).$$

Here $\varepsilon_{ij}$ is an antisymmetric tensor and $S_{ij}$ the "stretch tensor" that describe the sample imperfections. Because of a fundamental symmetry of the quantum Hall problem, "particle-hole" symmetry $\sigma_0(X) = \sigma_0(-X)$ and $\sigma_H(X) = 1-\sigma_H(-X)$, we are able to extract $\rho_0$ and $\rho_H$ from the measured resistivity tensor $\rho_{ij}$. Notably, for the PI transition this translates to $\rho_0(X) = 1/\rho_0(X)$ and a quantized value $\rho_H = 1$ (we here work in units $h/e^2$). Notice that for the PP transitions the steps in $\rho_H$ complicate the problem considerably [9].

The scaling results obtained by our analysis of the magnetotransport data taken on the InGaAs/InP HS #2 are [1,8]:

$$\rho_0(X) = e^{-X - \gamma X^3 - O(X^5)} \qquad (4),$$

$$\rho_H(X,\eta) = 1 + \eta(T)\rho_0(X) \quad ; \quad \eta(T) = \left(\frac{T}{T_1}\right)^{y_\sigma} \qquad (5).$$

Eq.4 describes the exponential dependence of the $\rho_0$ isotherms. By plotting $\ln\rho_0$ versus $X$, $1/\nu_0(T)$ is determined and subsequently the critical exponent $\kappa$ and the temperature $T_0$ are extracted from a log-log plot of $\nu_0(T)$ versus $T$. The third order correction term $-\gamma X^3$ is small in the region of interest [1]. The amplitude of Eq.4 is given by $\rho_0(B_c) = 1$ (in units of $h/e^2$).

Eq.5 describes the corrections to quantization of the Hall resistivity $\rho_H = 1$ at higher $T$. The correction term $\eta(T)\rho_0(X)$ indicates that under "ordinary" quantum Hall conditions ($X \neq 0$) the corrections are exponential in $T$, while at the critical point ($X = 0$) they are algebraic in $T$. The parameter $y_\sigma$ is the leading irrelevant exponent and $T_1$ indicates a cross-over temperature for scaling.

The function $\eta(T)$ can be determined by factoring out $\rho_0(X)$ from Eq.5 (see Ref.1). In this paper we do not analyze the Hall data further, but rather concentrate on the analysis of the longitudinal resistivity.

## 3. Sample choice and magnetotransport experiments

The magnetotransport experiments were carried out on two different low-mobility ($\mu$ = 16000-34000 cm$^2$/Vs) semiconductor structures: (i) In$_{0.43}$Ga$_{0.57}$As/InP HS's [1,6,8,10] and (ii) an In$_{0.2}$Ga$_{0.8}$As/GaAs QW [10,11]. In these structures the two-dimensional electron gas is located in the InGaAs alloy and the carriers undergo short-range potential scattering, which is a prerequisite for probing the quantum critical behaviour in the quantum Hall effect over a wide temperature range. In our experiments the electron density $n_e$ varies from 1.1 to 3.4 x10$^{11}$ cm$^{-2}$ (see Table 1). In the case of the InGaAs/GaAs QW an infrared led was used to tune the sample to the desired carrier concentration. The plateau-insulator phase transition takes place at a critical field $B_c$ = 7.5 T for the lowest density (InGaAs/GaAs QW) and at $B_c$= 26.4 T for the highest density (InGaAs/InP HS #2).

The longitudinal $R_{xx}$ and transverse $R_{xy}$ resistance were measured on samples in Hall bar geometry (length $W$ and width $L$) and the longitudinal and Hall resistivity are calculated by $\rho_{xx} = (W/L)R_{xx}$ and $\rho_{xy} = R_{xy}$, respectively. The experiments were carried out with a low frequency (2.6-13 Hz) lock-in technique and an excitation current of 1-5 nA. Special care was taken to reduce the phase change of the ac-signal at large resistance values. This allowed us to measure resistance values in the insulating phase up to ~ 1 MΩ. The experiments on the InGaAs/InP HS's were carried out at the HMFL in Nijmegen for $T$ = 0.1-4.5 K. Experiments on the InGaAs/GaAs QW at a carrier density $n_e$= 1.1x10$^{11}$ cm$^{-2}$ were performed at the WZI in Amsterdam in the temperature range 0.08-4.5 K, while experiments at higher densities $n_e$= 1.5 and 2.0 x10$^{11}$ cm$^{-2}$, were performed at the HMFL in Grenoble in the temperature range 0.08-1.2 K.

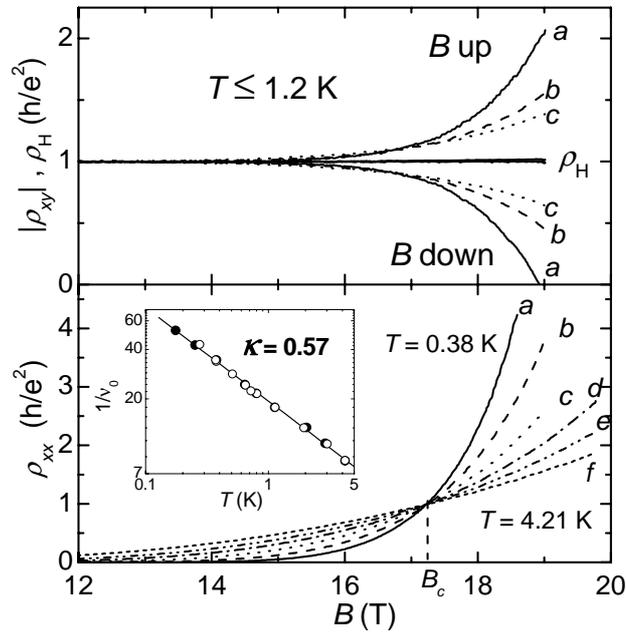

**Figure 1.** Longitudinal (lower frame) and Hall resistivity (upper frame) for the InGaAs/InP heterostructure #2 measured for opposite field directions as indicated. The Hall resistivity $\rho_H$ obtained by averaging over both field polarities is quantized at $h/e^2$ for $T \leq 1.2$ K. The letters $a,b,...f$ indicate $T$ = 0.38, 0.65, 1.2, 2.1, 2.9 and 4.2 K. Inset: double log plot of $1/\nu_0$ vs. $T$ at the PI transition. The closed and open symbols indicate values obtained for opposite directions of the magnetic field. Figure taken from Ref.1.

## 4. Results

In Fig.1 we show typical magnetotransport data, $\rho_{xx}(B)$ and $\rho_{xy}(B)$, taken on the InGaAs/InP HS #2 for temperatures $T \leq 4.2$ K [1,6,8]. The $\rho_{xx}$ isotherms show a fixed point at $B_c = 17.2$ T ($\nu_c = 0.55$) and the critical resistivity $\rho_{xx,c}$ is equal to $h/e^2$ within the uncertainty in the geometrical factor ($W/L = 0.61 \pm 0.04$). When the magnetic field polarity is reversed the data are identical, $\rho_{xx}(B) = \rho_{xx}(-B)$, as it should, and we conclude $\rho_{xx} \approx \rho_0$. In the region of the PI transition the data follow an exponential dependence $\rho_0(\nu,T) \propto \exp(-\Delta\nu/\nu_0(T))$ (Eq.4) and the slope of $\log\rho_0$ versus $\Delta\nu$ determines $\nu_0(T)$. In the inset in Fig.1 we show $\nu_0$ versus $T$ extracted for both magnetic field polarities on a log-log plot. The data follow a straight line over the entire temperature range, from which we determine the critical scaling exponent $\kappa = 0.57$, as well as $T_0 = 188$ K. In the upper panel of Fig.1 we have plotted the $|\rho_{xy}(B)|$ data as a function of $|B|$. This shows that a significant component of $\rho_{xx}$ is mixed into $\rho_{xy}$. Clean Hall data are obtained after averaging over the different field polarities, $\rho_{xy} = (\rho_{xy}(B\uparrow) + \rho_{xy}(B\downarrow))/2$. For $T \leq 1.2$ K, $|\rho_{xy}(B\uparrow)|$ and $|\rho_{xy}(B\downarrow)|$ are almost symmetric around the $i = 1$ plateau value and consequently $\rho_{xy} \approx h/e^2$ up to 19 T. For $T > 1.2$ K $|\rho_{xy}(B\uparrow)|$ and $|\rho_{xy}(B\downarrow)|$ are no longer symmetric around the value $h/e^2$ (see Refs.8,13) and as a result $\rho_{xy}$ deviates from the quantized value. The deviations become more pronounced upon raising the temperature and give access to the second critical exponent $y_\sigma$ (see Eq.5), as discussed in Refs.1,8,13.

The magnetotransport data of a second InGaAs/InP HS (#59) are reported in Ref.10. The experiments were performed in magnetic fields up to 30 T. In Fig.2 we show the isotherms $\rho_{xx}$ as a function of $\Delta\nu = \nu - \nu_c$ ($\nu_c = 0.53$) measured in the temperature range 0.14-4.5 K. For this sample the PI transition takes place at $B_c = 26.4$ T, as indicated by the fixed point for $T \leq 3.4$ K in Fig.2. At the critical point $\rho_{xx,c} = 1.08\ h/e^2 \approx \rho_0$, within the accuracy of the geometrical factor ($W/L = 0.37 \pm 0.03$). The data follow on exponential dependence $\rho_0(\nu,T) \propto \exp(-\Delta\nu/\nu_0(T))$ in the vicinity of the PI transition. Away from $\nu_c$ the deviations become significant, which indicates that in this region higher order terms in the scaling parameter $X$ should be taken into account (see Eq.4) to fit the data. In the inset of Fig.2 we show $\nu_0(T)$ versus $T$ on a log-log plot. Strictly speaking the data do not fall on a straight line. At the high temperature side this is due to thermal broadening by the Fermi-Dirac distribution. When fitting the data below 2 K, we obtain a critical exponent $\kappa = 0.58$ and $T_0 = 230$ K.

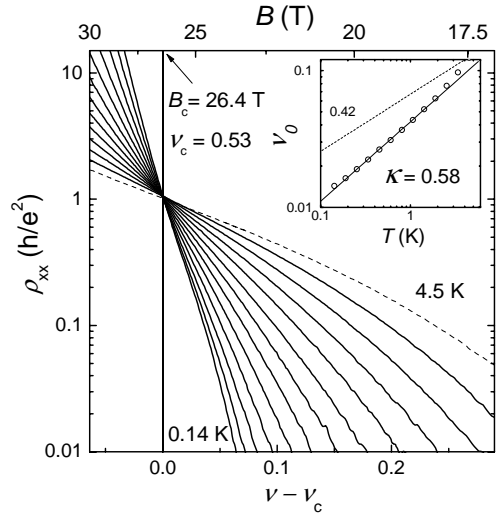

**Figure 2.** The longitudinal resistivity of the InGaAs heterostructure #59 near the PI transition as function of the filling factor (lower axis) or magnetic field (upper axis) at temperatures of 0.14, 0.19, 0.25, 0.34, 0.45, 0.6, 0.8, 1.1, 1.4, 1.9, 2.5, 3.4 and 4.5 K (dashed line). The inset is a double log plot of $\nu_0$ vs. $T$. The solid line gives $\kappa = 0.58$. For comparison we also show $\kappa = 0.42$ (dashed line). Figure taken from Ref.10.

In order to investigate the universality of our scaling results for the PI transition we have also conducted experiments on a quite different semiconductor structure, namely a low-mobility InGaAs/GaAs QW (well width 12 nm) [10,11]. The sample is insulating in the dark state, but could be tuned to different carrier densities by the persistent photoconductivity effect. In Fig.3 we present the main results for the longitudinal resistivity at densities $n_e$= 1.1 and 2.0 x$10^{11}$ cm$^{-2}$ in a plot of log$\rho_{xx}$ versus $\nu$. The data show clear fixed points at the critical filling factors $\nu_c$ = 0.63 and 0.53 ($B_c$ = 7.6 T and 15.7 T) for the different densities. The value of $\rho_{xx,c}$ equals $h/e^2$ to within 1% (i.e. the error in the ratio $W/L$= 0.19) for the highest density. For the lowest density $\rho_{xx,c}$ is about 10% smaller as indicated in Fig.3b. Fig.4 shows $\nu_0(T)$ versus $T$ on a log-log plot for $n_e$= 1.1, 1.5 and 2.0 x$10^{11}$ cm$^{-2}$ from which we extract $\kappa$ values of 0.53, 0.54 and 0.58, respectively.

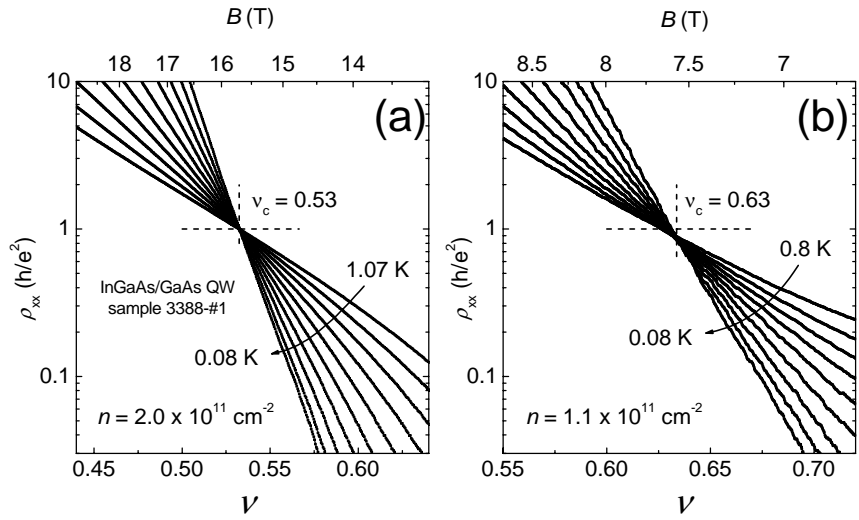

**Figure 3.** The longitudinal resistivity of the InGaAs/GaAS QW near the PI transition as function of the filling factor (lower axis) or magnetic field (upper axis) at temperatures of 0.08, 0.107, 0.14, 0.19, 0.26, 0.34, 0.45, 0.60, 0.80 and 1.07 K (in (a) only), for a carrier density $n_e$ equal to (a) 2.0 x$10^{11}$ cm$^{-2}$ and (b) 1.1x$10^{11}$ cm$^{-2}$. Data from Refs.10,11.

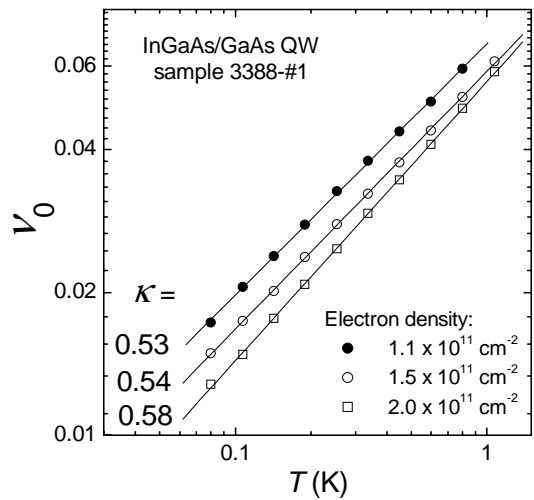

**Figure 4.** Double log plot of $\nu_0$ versus $T$ at the PI transition for the InGaAs/GaAs QW at three different densities as indicated. Data from Refs. 10,11.

**Table 1**. Results of the scaling analysis, i.e. the critical exponent $\kappa$ and the characteristic temperature scale $T_0$, of the longitudinal resistivity of various semiconductor structures with electron density $n_e$ and critical filling factor $\nu_c$. HS= heterostructure. QW=quantum well.

| Sample | $n_e$ ($10^{11}$ cm$^{-2}$) | $\nu_c$ | $\kappa$ | $T_0$ (K) | Ref. |
|---|---|---|---|---|---|
| InGaAs/InP HS #2 | 2.2 | 0.55 | 0.55 | 215 | [4] |
| - | 2.2 | 0.55 | 0.57 | 188 | [1],[6],[8] |
| InGaAs/InP HS #59 | 3.4 | 0.53 | 0.58 | 230 | [10] |
| InGaAs/GaAs QW | 1.1 | 0.63 | 0.53 | 160 | [10] |
| - | 1.5 | 0.58 | 0.54 | 185 | [10],[11] |
| - | 2.0 | 0.53 | 0.58 | 135 | [10],[11] |

## 5. Discussion and summary

In Table 1 we have collected the main results of the scaling analysis of the longitudinal resistivity of the investigated InGaAs/InP HS's and InGaAs/GaAs QW as presented in Figs 1-4. We have also included the results of a previous magnetotransport study on the InGaAs/InP HS #2 (Ref.4), which yielded $\kappa = 0.55$. The uncertainty in the values of $\kappa$ amounts to ±0.02, except in Ref.4 the error equals ±0.05. From the data in Table 1 we calculate a mean value $\kappa = 0.56 \pm 0.02$. The sample dependent characteristic temperature $T_0$, which indicates the full width of the Landau level, ranges from 135-230 K. The error in $T_0$ amounts to 10%. At the PI transition the critical resistivity $\rho_{xx,c} \approx h/e^2$, except for the InGaAs/GaAs QW at the lowest density, where $\rho_{xx,c}$ is about 10% lower. Notice that for all data sets the critical filling factor $\nu_c$ is typically 10% larger than the ideal value $\nu_c = 0.5$ for the PI transition. This we attribute to the non-negligible overlap of the lowest Landau level N=0↑ with the higher Landau level N=0↓. The overlap decreases with increasing density. Since the overlap is with *localized* states of the higher Landau level, it does not affect the transport properties.

The double log plots of $\nu_0(T)$ *versus* $T$ provide a clear demonstration of scaling, albeit over a somewhat limited temperature range: about one and a half decade in temperature for the InGaAs/InP sample #2 and about one decade in temperature for the other samples. At the high temperature side the observation of scaling is hampered by the thermal broadening due to the Fermi-Dirac distribution, while at the low-temperature side macroscopic sample inhomogeneities (notably carrier gradients) limit the proper analysis with help of Eq.3 to $T > 0.1$ K. We conclude that our $\rho_{xx}$ data at the PI transition obey the scaling theory of the quantum Hall effect with a critical exponent $\kappa = 0.56 \pm 0.02$.

The critical exponent observed for the PI transition is larger than the value $\kappa = 0.42 \pm 0.05$ [14] (see dashed line in the inset of Fig.2) repeatedly reported in the literature for the PP transition. At this stage it is important to stress that our InGaAs/InP samples #59 and #2 are the very same InGaAs/InP Hall bars as measured in the seminal papers by Wei *et al.* [14] and Hwang *et al.* [15], respectively. In Ref. 14 the quantum Hall scaling law for the PP transitions in the Landau levels N=0↓, N=1↑ and N=1↓ was demonstrated for the first time. The maximum slope in the Hall resistivity $(d\rho_{xy}/dB)_{max}$ and the inverse of the half width of $\rho_{xx}$ between adjacent quantum Hall plateaus $(\Delta B)^{-1}$ were both found to diverge algebraically with temperature as $T^{-\kappa}$ with a critical exponent $\kappa = 0.42$. In Ref.15 sample #2 was used to investigate spin polarized *and* spin degenerate levels by tilting the sample with respect to the magnetic field. The critical exponents for the spin degenerate levels were found to be a factor two smaller than those for the spin polarized levels $\kappa = 0.42$. Thus, while $\kappa = 0.42$ is observed in our InGaAs/InP samples for the spin polarized PP transitions, the critical exponent for the PI transition equals $\kappa = 0.56$. However, within the scaling theory for the integral quantum Hall effect the spin polarized PP and PI transitions should have the same critical exponents [16]. This solution for this discrepancy lies in understanding and modeling the influence of macroscopic sample inhomogeneities on the transport tensor at the PP and PI transitions [1,8,9]. An important result of this work is that these inhomogeneities, when small, do not thwart a proper determination of $\kappa$ for the PI transition. In other words, the PI transition is robust against sample imperfections. However, sample imperfections considerably complicate the analysis for the PP transitions.

Recently, we have been able to estimate the influence of small carrier density gradients on the value of the critical exponent [10]. Numerical simulations of the quantum Hall transport problem in Hall bar geometry show that for the PI transition a carrier gradient of a few percent along the Hall bar does not affect the critical exponent $\kappa$. However, for the 2→1 PP transition the situation is very different and a considerably smaller "effective" $\kappa$-value is found. The reduction of $\kappa$ is typically of the order of 10% for a simulated density gradient of 2% along the Hall bar. Such density gradients are in-line with the experimental observations. A density gradient along the Hall bar of 2-3 % has been measured for our InGaAs/InP heterostructure #2 [1], while a detailed investigation of the resistivity tensor using different contact pairs of the Hall bar of the InGaAs/InP heterostructure #59 revealed density variations of ~3% [10]. Thus the presence of macroscopic sample inhomogeneities in our samples may indeed explain the smaller $\kappa$-values for the PP transitions. On the other hand, recent state of the art experiments [17] on $Al_xGa_{1-x}As/Al_{0.33}Ga_{0.67}As$ quantum wells with controlled short-ranged alloy potential fluctuations confirmed universal scaling with $\kappa = 0.42$ for PP transitions in higher Landau levels ($N = 1\downarrow$ and higher) in the range $0.0065 < x < 0.016$. Interestingly for larger values of $x$, $\kappa$ increases to ~0.58, which was attributed to the break-down of universal scaling due to clustering of Al atoms.

In the framework of the scaling theory of the quantum Hall effect [16] the critical exponent $\kappa = p/2\chi$, where $p$ is the exponent of the phase breaking length $l_\varphi$ at finite temperature (i.e. $l_\varphi \sim T^{-p/2}$) and $\chi$ the critical exponent for the zero $T$ localization length $\xi$. In the literature often the experimental estimate $\chi \sim 2.4$ is quoted, which is deduced from $\kappa = 0.42$ for the PP transitions [14] and $p = 2$ obtained by current scaling experiments [18]. This value is close to the Fermi liquid value 7/3 obtained by numerical simulations (see e.g. Ref.19). However, this Fermi liquid scenario has always been quite confusing as in a microscopic theory of the quantum Hall effect Coulomb interactions should be incorporated. With our new value $\kappa = 0.56$ and $1 < p < 2$ [20], we calculate that the correlation length exponent is bounded by $0.9 < \chi < 1.8$. This indicates a non-Fermi liquid universality class for the integral quantum Hall transitions.

In summary, we have presented a review of recent low-temperature high-field magnetotransport studies of the plateau-insulator transition in several InGaAs/InP heterostructures and an InGaAs/GaAs quantum well. Our sample choice is motivated by the uncorrelated δ-function-like potential fluctuations present in the InGaAs alloy. The InGaAs/InP samples investigated in this work are the same Hall bars as previously used in the pioneering scaling experiments on the plateau-plateau transitions [14,15]. We find that the longitudinal resistivity $\rho_{xx}$ near the critical filling factor for the plateau-insulator transition $\nu_c \approx 0.5$ follows the universal scaling law $\rho_{xx}(\nu,T) \propto \exp(-\Delta\nu/(T/T_0)^\kappa)$, where $\Delta\nu = \nu-\nu_c$ with a critical exponent $\kappa = 0.56\pm0.02$. The scaling behaviour is observed over roughly one decade in temperature. The low temperature limit for scaling in our experiments is set by macroscopic sample inhomogeneities, which hamper the extraction of the intrinsic resistivities from the experimental data below $T \sim 0.1$ K. The new exponent $\kappa = 0.56\pm0.02$ is larger than the "accepted" exponent $\kappa = 0.42$ for the PP transitions. Numerical simulations of the quantum Hall transport problem [10] show that the smaller value observed for the PP transitions can be explained by a small (2-3%) carrier density gradient along the Hall bar, which broadens the transition and mask the true critical behaviour.

**Acknowledgements** This work was part of the research programme of FOM (Dutch Foundation for Fundamental Research of Matter).